\documentclass[prd,twocolumn,showpacs]{revtex4}

\usepackage{amsmath,amssymb,epsfig,graphics}

\def\be{\begin{equation}}
\def\ee{\end{equation}}

\def\d{{\rm d}}
\def\e{{\rm e}}
\def\s{{\rm sech}(w(\tau-\tau_0))}
\def\t{{\rm tanh}(w(\tau-\tau_0))}

\def\sN{\mathcal{N}}
\def\EEE{E_1{}^1}
\def\Sm{\Sigma_-}
\def\Sc{\Sigma_\times}
\def\St{\Sigma_2}
\def\Nm{N_-}
\def\Nc{N_\times}

\def\Sp{\Sigma_+}
\def\Np{N_+}
\def\lap{-\tfrac12}
\def\dt{\partial_\tau}
\def\dx{\partial_x}
\def\dT{\partial_T}
\def\dX{\partial_X}

\def\Sph{\Sigma_+^H}
\def\Smh{\Sigma_-^H}

\def\sech{\, {\rm sech}\, }

\def\di{e^{AT}\EEE\partial_X}
\def\E{\mathcal{E}}
\def\H{\mathcal{H}}
\def\Em{\E_-}
\def\Ec{\E_\times}
\def\Hm{\H_-}
\def\Hc{\H_\times}
\def\Ep{\E_+}
\def\Hp{\H_+}

\begin{document}

\title{Spikes in the Mixmaster regime of $G_2$ cosmologies}

\author{Woei Chet Lim}
\email{wclim@aei.mpg.de}
\affiliation{Albert-Einstein-Institut, Am M{\"u}hlenberg 1, D-14476 Potsdam, Germany.}
\affiliation{Joseph Henry Laboratories, Princeton University, Princeton, NJ 08544, USA.}

\author{Lars Andersson}
\email{lars.andersson@aei.mpg.de}
\affiliation{Albert-Einstein-Institut, Am M{\"u}hlenberg 1, D-14476 Potsdam, Germany.}
\affiliation{Department of Mathematics, University of Miami, Coral Gables, FL 33124, USA.}

\author{David Garfinkle}
\email{garfinkl@oakland.edu}
\affiliation{Department of Physics, Oakland University, Rochester, Michigan 48309, USA.}

\author{Frans Pretorius}
\email{fpretori@princeton.edu}
\affiliation{Joseph Henry Laboratories, Princeton University, Princeton, NJ 08544, USA.}

\date{\today}

\begin{abstract}
We produce numerical evidence that spikes in the Mixmaster regime of $G_2$ 
cosmologies are transient and recurring, supporting the conjecture that
the generalized Mixmaster behavior is asymptotically non-local where 
spikes occur.
Higher order spike transitions are observed to split into separate first order spike
transitions.

\

\noindent
AEI preprint number: AEI-2009-033
\end{abstract}
\maketitle

\section{Introduction}

Belinskii, Khalatnikov and Lifshitz (BKL) \cite{LK63,BKL70,BKL82} 
conjectured that, according to general relativity, the approach to the 
generic spacelike singularity is vacuum dominated (assuming $p<\rho$), 
local, and oscillatory (labeled `Mixmaster'). 
Here local means that the contribution of terms in the evolution
equations with spatial derivatives becomes negligible.  
The Mixmaster dynamics consists of Kasner epochs bridged by transitions
represented by the vacuum Bianchi type II solution. 
Numerical studies of the asymptotics of the Gowdy models, which 
represent the 
simplest inhomogeneous vacuum spacetimes, reveal that on 
approach to the singularity, spiky structures form
\cite{BM93,BG98,HS98}.  These spikes become ever narrower as the 
singularity is approached.  At first, the presence of such spikes might
seem inconsistent with the 'local' part of the BKL conjecture since
the spatial derivative of such a spike grows without bound as the
singularity is approached.  Remarkably, such spiky behavior in Gowdy
spacetimes actually {\it is} consistent with BKL locality.  The 
reason for this is that in the evolution equations for Gowdy spacetimes,
the spatial derivatives are multiplied by a quantity that goes to zero
even faster than the spatial derivatives go to infinity.  This has been
verified in detailed numerical simulations, as well as by the discovery 
of closed form Gowdy solutions with the spike property \cite{RW01,L08}.  

Nonetheless, the Gowdy models are a very special class of spacetimes,
so it remains to be seen whether this property of BKL locality and 
persistent spikes holds in more general spacetimes.  To that end, we
will examine the properties of spikes in $G_2$ models which are a slightly
more general class that includes the Gowdy spacetimes. 

Studies of $G_2$ and more general models have produced 
numerical evidence that the BKL conjecture generally holds except possibly 
at isolated points where spiky structures  
form \cite{H99,BIW01,G04,AELU05}. 
Here, BKL locality is violated due to large spatial gradients. 
However, the ability to draw conclusions about spikes from such
simulations is severely limited 
due to the enormous numerical resources needed to resolve the narrowing 
spikes. In this paper, we will use a different numerical method that 
does have adequate resolution to provide reliable conclusions about 
spike behavior in $G_2$ spacetimes. 
We present numerical evidence in support of the following conjectures:
that recurring ``spike transitions" are a general type of oscillation as the singularity is approached,
and that higher-order spike transitions split into separate first-order spike transitions.
 Section 2 presents the equations 
for the evolution of $G_2$ spacetimes.  Our numerical method is presented in 
section 3, results in section 4, and conclusions in section 5.
Appendix~\ref{App_A} presents the procedure to match a numerical solution with an explicit spike solution.
Appendix~\ref{App_B} gives the formula for the BKL parameter $u$ in term of the parameter $w$.
Appendix~\ref{App_C} gives the formulae for the Weyl scalar invariants. 

\section{$G_2$ spacetimes}

The metric of the general $G_2$ class takes the form
\cite[eq (7)]{BIW01}
\begin{align}
\d s^2 &= - \e^{(\lambda-3\tau)/2} \d\tau^2 
	 + \e^{(\lambda+\mu+\tau)/2} \d x^2
\notag\\
         &\quad+ \e^{P-\tau} [ \d y+Q\, \d z 
	   + (G_1 +Q G_2) \d x 
\notag\\
	&\qquad\qquad\quad + (M_1 + Q M_2)(-\e^{-\tau} \d \tau)]^2 
\notag\\
	 &\quad+ \e^{-P-\tau} [\d z +G_2 \d x + M_2 (-\e^{-\tau}\d 
\tau)]^2.
\end{align}
Here all metric quantities depend only on the time coordinate $\tau$
and spatial coordinate $x$, thus there is symmetry in two spatial
directions.  The singularity is approached as $\tau \to \infty$.
The choice of gauge used here is the same as in \cite{BIW01,AELU05}.

Our choice of variables are the $\beta$-normalized variables 
\cite{vEUW02,thesis} in the orthonormal frame formalism \cite{vEU97}, 
related to the metric components as follows:
\begin{gather}
        \beta = 2\e^\frac{\lambda-3\tau}{4},\qquad
	\sN = -\frac12,
\\
        \EEE = 2 \e^{-\frac{\mu}{4}-\tau},\qquad
        \St = \frac{K}{\sqrt{3}} \e^\frac{\lambda+2P+3\tau}{4},\qquad
\\
	\Sm = -\frac{P_\tau}{\sqrt{3}},\qquad
	\Nc = -\frac{\e^{-\frac{\mu}{4}-\tau}P_x}{\sqrt{3}},
\\
	\Sc = -\frac{\e^P Q_\tau}{\sqrt{3}},\qquad
	\Nm =  \frac{\e^{P-\frac{\mu}{4}-\tau} Q_x}{\sqrt{3}},
\end{gather}
where $K$ is a constant, and the $\tau$ and $x$ subscripts denote partial differentiation.

The evolution equations for the $\beta$-normalized variables are:
\begin{align}
	\dt\EEE &= \lap (2-3\St^2)\EEE
\label{dte11}
\\
	\dt\Sm &= \lap[ -3\St^2 \Sm +2\sqrt{3}(\Sc^2-\Nm^2) -\sqrt{3}\St^2
\notag\\
	&\qquad\quad
			-\EEE \dx\Nc ]
\\
	\dt\Nc &= \lap[ (2-3\St^2)\Nc - \EEE \dx\Sm ]
\\
	\dt\Sc &= \lap[ (-3\St^2-2\sqrt{3}\Sm)\Sc -2\sqrt{3}\Nc\Nm 
\notag\\
        &\qquad\quad
			+ \EEE \dx\Nm ]
\\
	\dt\Nm &= \lap[ (2-3\St^2+2\sqrt{3}\Sm)\Nm + 2\sqrt{3}\Sc\Nc
\notag\\
        &\qquad\quad
			+ \EEE \dx\Sc ]
\\
	\dt\St &= \lap [-3\St^2 -3\Sp +\sqrt{3}\Sm]\St,
\end{align}
where
\be
	\Sp = \frac12(1-\Sm^2-\Sc^2-\St^2-\Nm^2-\Nc^2).
\ee
There is one constraint equation:
\be
	\EEE \dx\St = (3\Nm\Sc-3\Nc\Sm-\sqrt{3}\Nc)\St.
\ee
For state-space presentations, we will use the Hubble-normalized 
variables \cite{L08}:
\begin{multline}
        (\Sp,\Sm,\Sc,\St,\Nm,\Nc)^H
\\
 = \frac{1}{1-\Sp}(\Sp,\Sm,\Sc,\St,\Nm,\Nc).
\end{multline}
See \cite{AELU05} for the evolution equations for Hubble-normalized 
variables, and the derivation of the evolution equations.

The Gowdy spacetimes are that class of $G_2$ spacetimes for
which ${\Sigma_2}=0$.  Note that it then follows from equation 
(\ref{dte11}) that
${{E_1}^1} = \exp (-\tau )$.  An interesting class of solutions 
of the Gowdy equations are the exact spike solutions of \cite{L08}  
\begin{align}
\Sm &= {\frac {-1} {\sqrt 3}} \bigl ( 1 + {\frac {{f^2} - 1} {{f^2}+1}}
[ w \tanh (w\tau ) - 1 ] \bigr )
\\
\Nc &= {\frac {2 f} {{f^2}+1}} {\frac w {\sqrt 3}} \sech (w \tau)
\\
\Sc &= {\frac {{f^2} - 1} {{f^2}+1}} {\frac w {\sqrt 3}} \sech (w \tau)
\\
\Nm &=  {\frac {2 f} {{f^2}+1}} {\frac 1 {\sqrt 3}} ( 1 - w \tanh (w\tau ))
\end{align}
where $w$ is a constant and the quantity $f$ is given by
\be
f = w {e^\tau} \sech (w\tau ) x
\ee
For $|w|<1$ this solution describes a spike because $f=0$ at $x=0$ but
$f$ becomes large as $\tau \to \infty$ for all $x \ne 0$.  Nonetheless,
BKL locality is preserved
because in the equations of motion
all spatial derivatives are multiplied by ${{E_1}^1}$  
and ${{E_1}^1} {\partial _x} f = w \sech (w \tau)$ which goes to 
zero as $\tau \to \infty$. 

Note however that this conclusion depends on the fact that 
$ {{E_1}^1} = \exp (-\tau)$ which in turn depends on the fact 
that in equation (\ref{dte11}) we could set $\St$ to zero, something
that we can only do in Gowdy spacetimes, not the more general $G_2$
spacetimes.  The dynamics in a $G_2$ spacetime consists of eras where
$\St$ is very small (and which can thus be well described by the dynamics
of Gowdy spacetimes) punctuated by short "frame bounces" where $\St$ 
rapidly grows
and then rapidly shrinks to become again negligible.  During a frame
bounce ${E_1}^1$ shrinks more slowly than $\exp (-\tau)$ and thus it 
is not clear whether spatial derivatives continue to remain negligible.
To resolve this issue, we will need to perform numerical simulations of
the dynamics of $G_2$ spacetimes. Furthermore, those simulations will 
need to have enough resolution to accurately model the rapidly shrinking
spikes.   

\section{Numerical methods}

One numerical method for resolving small scale structure is adaptive 
mesh refinement (AMR).  However, if one knows beforehand the location
of the structure, one need not use AMR and can instead use a coordinate
system adapted to the structure that one wants to study.  In particular,
here we are studying spikes that shrink exponentially with time, so we
choose a coordinate system that does the same.

\begin{figure}[t!]
    \resizebox{\columnwidth}{!}{\includegraphics{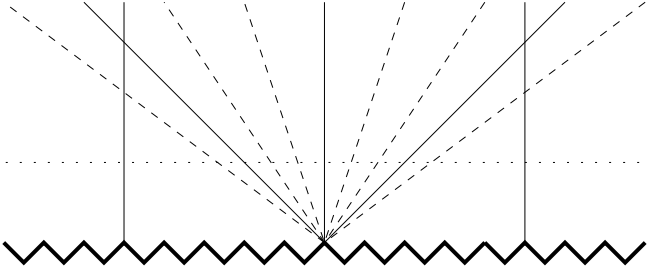}}
    \caption{\label{fig2}The spacetime diagram showing the $x=\text{const}$ worldlines 
(vertical), the spatial hypersurfaces $\tau=T=\text{const}$ (horizontal) 
with $\tau=T \rightarrow \infty$ at the singularity, the particle horizon 
($45^\circ$ lines) of the observer in the center, and the 
$X=\text{const}$ lines which are timelike inside the particle horizon 
but spacelike outside (dashed lines).} 
\end{figure}

We introduce new coordinates $(T,X)$ to zoom in on the 
worldline $x=x_\text{zoom}$.
\be
        T = \tau,\qquad X=\e^{A\tau} (x-x_\text{zoom}),
\ee
where the constant $A$ controls the rate of focus. See Figure~\ref{fig2} 
for the qualitative spacetime diagram.
The differential operators expressed in the new coordinates are
\be
        \dt = \dT + AX \dX,\qquad \dx =  \e^{AT} \dX.
\ee
The equations in the new coordinates are
\begin{align}
        \dT\EEE &= - AX \dX\EEE - [1-\tfrac32 \St^2]\EEE\\
        \dT\Sm &= - AX \dX\Sm
                + \tfrac12\e^{AT}\EEE \dX\Nc
                + \tfrac32 \St^2 \Sm
\notag\\
	&\qquad
		- \sqrt{3}(\Sc^2-\Nm^2)
                + \tfrac{\sqrt{3}}{2}\St^2
\\
        \dT\Nc &= - AX \dX\Nc
                + \tfrac12\e^{AT}\EEE \dX\Sm
                - [1-\tfrac32 \St^2 ] \Nc
\label{dT_Nc}
\\
        \dT\Sc &= - AX \dX\Sc
                - \tfrac12\e^{AT}\EEE \dX\Nm
                + \tfrac32 \St^2 \Sc
\notag\\
        &\qquad
                + \sqrt{3}\Sm\Sc + \sqrt{3}\Nc\Nm
\\
        \dT\Nm &= - AX \dX\Nm
                - \tfrac12\e^{AT}\EEE \dX\Sc
                - [1-\tfrac32 \St^2 ] \Nm
\notag\\
        &\qquad
                - \sqrt{3}\Sm\Nm - \sqrt{3}\Sc\Nc
\\
        \dT\St &= - AX \dX \St
        + [ \tfrac32 \St^2 +\tfrac32\Sp
                - {\tfrac {\sqrt{3}} 2} \Sm]\St
\label{conX}
\end{align}
and constraint
\be
        \e^{AT} \EEE \dX\St = (3\Nm\Sc-3\Nc\Sm-\sqrt{3}\Nc)\St.
\ee

We will end the numerical grid at a fixed coordinate value $X={X_0}$.  
Ordinarily,
that would call for a boundary condition at $X_0$, but we will 
use the method of excision.  Usually one thinks of excision as
applying to simulations of black holes; however excision can be
applied to any hyperbolic equations where the outer boundary
is chosen so that all modes are outgoing.  In that case one 
simply implements the equations of motion at the outer boundary, 
no boundary condition is needed (or even allowed).  

The following combinations of the equations of motion
\begin{align}
        \dT(\Sm+\Nc) &= -(AX-\tfrac12\e^{AT}\EEE) \dX(\Sm+\Nc) + \cdots
\\
        \dT(\Sm-\Nc) &= -(AX+\tfrac12\e^{AT}\EEE) \dX(\Sm-\Nc) + \cdots
\\
        \dT(\Sc-\Nm) &= -(AX-\tfrac12\e^{AT}\EEE) \dX(\Sc-\Nm) + \cdots
\\
        \dT(\Sc+\Nm) &= -(AX+\tfrac12\e^{AT}\EEE) \dX(\Sc+\Nm) + \cdots
\end{align}
clearly shows that $(\Sm+\Nc)$ and $(\Sc-\Nm)$ flow away from 
$X=\tfrac{1}{2A}\e^{AT}\EEE$ (for $A>0$), while
$(\Sm-\Nc)$ and $(\Sc+\Nm)$ flow away from $X=-\tfrac{1}{2A}\e^{AT}\EEE$. 
This puts $X=\pm\tfrac{1}{2A}\e^{AT}\EEE$ as the points beyond which the 
flow is entirely outward.
Thus, as long as $X_0$ is chosen large enough and as long as 
${e^{AT}} {{E_1}^1}$ does not grow too large during the simulation,
the surface $X = {X_0}$ will be a good excision boundary.

In addition to choosing $X_0$, we should also choose $A$ so that
$X={X_0}$ remains a good excision boundary throughout the simulation. 
$A=1$ is the natural choice,
which fixes the particle horizon of the exact spike solution as a vertical line
in the spacetime diagram with respect to ($X$,$T$) coordinates. In this paper we shall choose $A=1$.
Choosing another value for $A$ is a trial and error process, 
but one is able estimate $\EEE$ after one or two numerical runs, with the heuristics below.

We shall define phenomenologically that a Gowdy era as the time period during which $\St$ is small.
We take this opportunity to correct that the Kasner eras mentioned in \cite{L08} are in fact Gowdy eras.
The two are not equivalent, as there can be two or three Kasner eras within one Gowdy era.
During a Gowdy era, $\EEE$ approximately equals $e^{-\tau}$, but between Gowdy eras 
(namely during the $\St$ transition) $\EEE$ shrinks more slowly.
In order to offset this behavior between Gowdy eras,
one should choose a small enough $A<1$ so that 
$ {e^{A T}}{{E_1}^1}$ 
decays during the Gowdy era.
But if $A$ is too small, spikes will be inadequately resolved. 
A reasonable range is $0.8 \leq A \leq 1$.
 
Another way to make $X={X_0}$ a good excision boundary is to 
choose a larger $X_0$ to leave 
more room for the growth of 
$ {e^{A T}}{{E_1}^1}$.
The CFL condition, however, requires 
that the numerical time step $\Delta T$ satisfies
\be
	\Delta T < \left( A X_0+\frac12 e^{AT} \EEE \right)^{-1}
			\Delta X,
\ee
where $\Delta X$ is the numerical grid size.
For example, doubling $X_0$ would cut $\Delta T$ by a half, so one is 
bound by numerical resources to choose a large enough $X_0$ for the 
simulation without being too wasteful.
A reasonable range is $10 \leq X_0 \leq 40$.

Our numerical simulations use a uniform spatial grid.
The equations are evolved using the classical fourth-order Runge-Kutta 
method, with fourth-order accurate spatial derivatives.
That is for any quantity $F$ we approximate ${\partial _X}F$ on
grid point $i$ by
\be
\frac43 {\frac {{F_{i+1}}-{F_{i-1}}} {2 \Delta X}}
- \frac13 {\frac {{F_{i+2}}-{F_{i-2}}} {4 \Delta X}}
\ee
On the last gridpoint, the excision boundary, we evaluate the spatial
derivative using one sided differences.  That is we approximate
${\partial _X}F$ at the final gridpoint $N$ by
\be
\frac{25 F_N - 48 F_{N-1} + 36 F_{N-2} - 16 F_{N-3} + 3 F_{N-4}}{12 \Delta X}
\ee
and at the second last grid point $N-1$ by
\be
\frac{3 F_N + 10 F_{N-1} - 18 F_{N-2} + 6 F_{N-3} - F_{N-4}}{12 \Delta X}
\ee
For spikes, non-symmetric data would be problematic for implementing the
local perspective as the spike worldline is not stationary in this case.
Therefore we shall choose symmetric initial data (around $X=0$) and simulate 
only $X\in[0,X_0]$, with enforcement of the symmetry at the left 
boundary $X=0$. For comparison, we also simulate along non-spike worldlines,
in which case the data are not symmetric and the left boundary at $-X_0$
is an excision boundary.

We choose the first gridpoint to be either an excision boundary
at $X=-{X_0}$ or a point of symmetry at $X=0$.  If it is an excision
boundary, then ${\partial _X}F$ is approximated by the one sided 
differences
\be
\frac{- F_1 + 48 F_2 - 36 F_3 + 16 F_4 - 3 F_5}{12 \Delta X}
\ee
at the first grid point, and
\be
\frac{- 3 F_1 - 10 F_2 + 18 F_3 - 6 F_4 + F_5}{12 \Delta X}
\ee
at the second.
However, if first grid point is a point of symmetry then we choose all 
quantities to be either even or odd there.  For even functions, 
${\partial _X}F=0$ at the first grid point,
and
\be
\frac43 {\frac {{F_{3}}-{F_{1}}} {2 \Delta X}}
- \frac13 {\frac {{F_{4}}-{F_{2}}} {4 \Delta X}}
\ee
and the second,
while for odd functions we approximate
${\partial _X}F$ by
\be
\frac{8 F_2 - F_3}{6 \Delta X}
\ee
at the first grid point, and
\be
\frac43 {\frac {{F_{3}}-{F_{1}}} {2 \Delta X}}
- \frac13 {\frac {{F_{4}}+{F_{2}}} {4 \Delta X}}
\ee
at the second.

The standard double precision real variables (with 16 digits of significance) 
are normally used in the numerical code.
When necessary, quad precision real variables (with 32 digits of significance) are used
to lower the numerical roundoff errors by $10^{16}$ folds,
thereby preventing it from prematurely swamping small values.
The variables $\Nm$, $\Sc$ and $\Nc$ take small values during Kasner epochs, and can
be swamped by the roundoff error in the spatial derivative term of another variable with
a larger value. Usually this happens to $\Nc$ first, when the term 
$\tfrac12\e^{AT}\EEE \dX\Sm$ in equation (\ref{dT_Nc}) 
becomes $10^{16}$ times smaller (if double precision is used) than the value of $\Sm$.
The usage of quad precision real variables increases the runtime by 4 to 8 folds.

No numerical dissipation is used, as it is unnecessary.

To verify that numerical solutions converge with fourth order accuracy,
we compare the constraint (\ref{conX}) in numerical runs with different 
resolutions (different number of grid points). 
We observe that doubling the resolution reduces the 
constraint by a factor of 16 when adequate numerical resolution is used.
We also compare the numerical solutions with a matching exact spike solution.
The procedure for matching is described in Appendix~\ref{App_A}.
The formula for the BKL parameter $u$ for the Kasner epochs between transitions
are given in Appendix~\ref{App_B}. The Weyl scalar invariants are used to measure the
difference between numerical and exact solutions. Appendix~\ref{App_C} gives their
formulae.

In this paper, we shall focus on obtaining numerically accurate results, which require
much higher numerical resolution than qualitative numerical results do. 
This requirement also places severe limit on how far into the asymptotic regime
one can simulate, because the numerical error must not be larger than the distance 
from the solution to the nearest Kasner point in the state space, and this in turn
require high numerical resolution. 
When a numerical simulation takes up to months to run in order to meet the accuracy, 
it becomes impractical.
Despite this difficulty, we want to provide more than just qualitative numerical
results, because numerically accurate results can provide evidence supporting convergence
to the exact spike solution, while qualitative numerical results cannot. 
In presentation, we shall round the numbers to 4 decimals, even though the accuracy is higher.

Qualitative numerical results are still valuable in providing evidence
supporting the general behavior of the solution.
Compared with other aspects of the solution, the timing of a transition is most sensitive to numerical inaccuracy.
At lower resolutions, the timing of a transition differs greatly while other aspects of the solution remain robust.

\section{Results}

We clarify a few terms we use below.
A (true) spike point is where a (true) spike can occur (the spike may be 
active or smoothed out). 
In our variables, a spike point is where 
\be
	\Nm=0.
\ee
We shall hold the spike point fixed (at $x=0$), so that we 
can easily locate it and zoom in on it. To do so, we require
$\Nm$ and $\Nc$ to be odd functions around the spike point, and
$\Sm$, $\Sc$ and $\St$ to be even functions.

A false spike point is where $\Sc=0$. To hold it fixed, we require 
$\Sc$ and $\Nc$ to be odd functions around the false spike point, and 
$\Sm$, $\St$ and $\Nm$ to be even functions.

We will present three sets of numerical results. 
The first set chooses a perturbed spike solution as the initial condition,
and shows two recurrences of the spike solution, within the same Gowdy era. 
The purpose is to show that spike recurs within the same Gowdy era.
The second set chooses a generic initial condition, and shows two occurrences of the spike solution,
one in each Gowdy era. 
The purpose is to show that spike recurs over different Gowdy eras.
The third set consists of two simulations, with a perturbed second and third order spike solution as the initial condition, respectively.
The purpose is to show that second and third order spikes break up into separate first order spikes.
All three sets demonstrate the attractor nature of the first order spike.

The reader will notice that different numerical resolutions are used for different sets. The length of simulation also differs.
Both numerical resolutions and length are not arbitrarily chosen, but are dictated by the cost of computation to maintain accuracy.
For example, in the first set, we do not try to extend the simulation to show the third spike recurrence over different Gowdy eras,
as it would be too costly.

\subsection{Perturbed spike}

The format of the initial data is a perturbed spike solution at $\tau=0$:
\begin{gather}
\EEE=2,\qquad \St = 10^{-5},
\\
\Sm = \frac{(wx)^2-1}{(wx)^2+1}\frac{1}{\sqrt{3}}
	-\frac{1}{\sqrt{3}} + \epsilon ,\qquad
\Nc = \frac{2wx}{(wx)^2+1} \frac{w}{\sqrt{3}},
\label{ic1b}
\\
\Sc = \frac{(wx)^2-1}{(wx)^2+1} \frac{w}{\sqrt{3}},\qquad
\Nm = \frac{2wx}{(wx)^2+1} \frac{1}{\sqrt{3}},
\label{ic1c}
\end{gather}
where $x=X+x_\text{zoom}$,
with $\epsilon=0.001$, $w=9.5$, $A=1$, $x_\text{zoom}=0$.
Here, small perturbations are applied to the variables $\St$ and $\Sm$.
Compare with the exact spike solution at $\tau=0$ \cite[eq (36)]{L08}.
The value $w=9.5$ allows for two spike recurrences at roughly
$w=5.5$ and $w=1.5$ before the next Gowdy era.

Because the data chosen are symmetric about $X=0$, only $X\geq0$ needs to
be simulated.
A resolution of 100001 grid points on the $X$-interval [0,2] is
sufficient for convergence during the time interval $T \in [0,16]$.
Double precision is used.
Beyond $T=16$ a higher numerical precision is needed to maintain accuracy.
To compare the orbit along a different worldline, we also use the 
same initial data but with $x_\text{zoom}=1$ 
and 200001 grid points on the $X$-interval $[-2,2]$.

\begin{figure*}[t!]
    \resizebox{\linewidth}{!}{\includegraphics{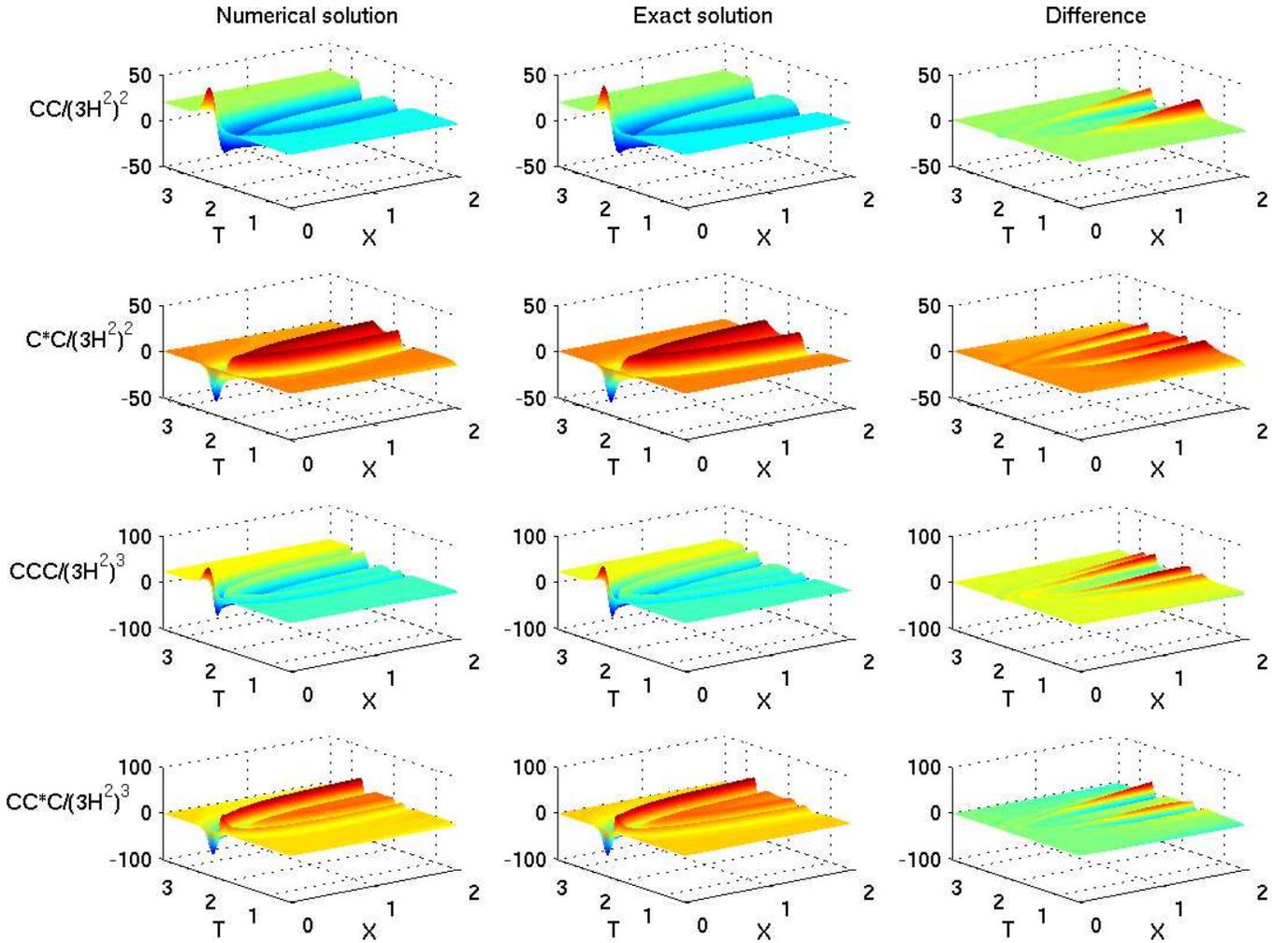}}
    \caption{\label{weyl_1}The Weyl scalar invariants (normalized by Hubble scalar)
    for the numerical solution, the matching exact spike solution (with $w=5.8644$, $\tau_0=2.2205$), 
    and their difference, during the first spike recurrence over the time interval $[0.5,3.5]$.} 
\end{figure*}

\begin{figure*}[t!]
    \resizebox{\linewidth}{!}{\includegraphics{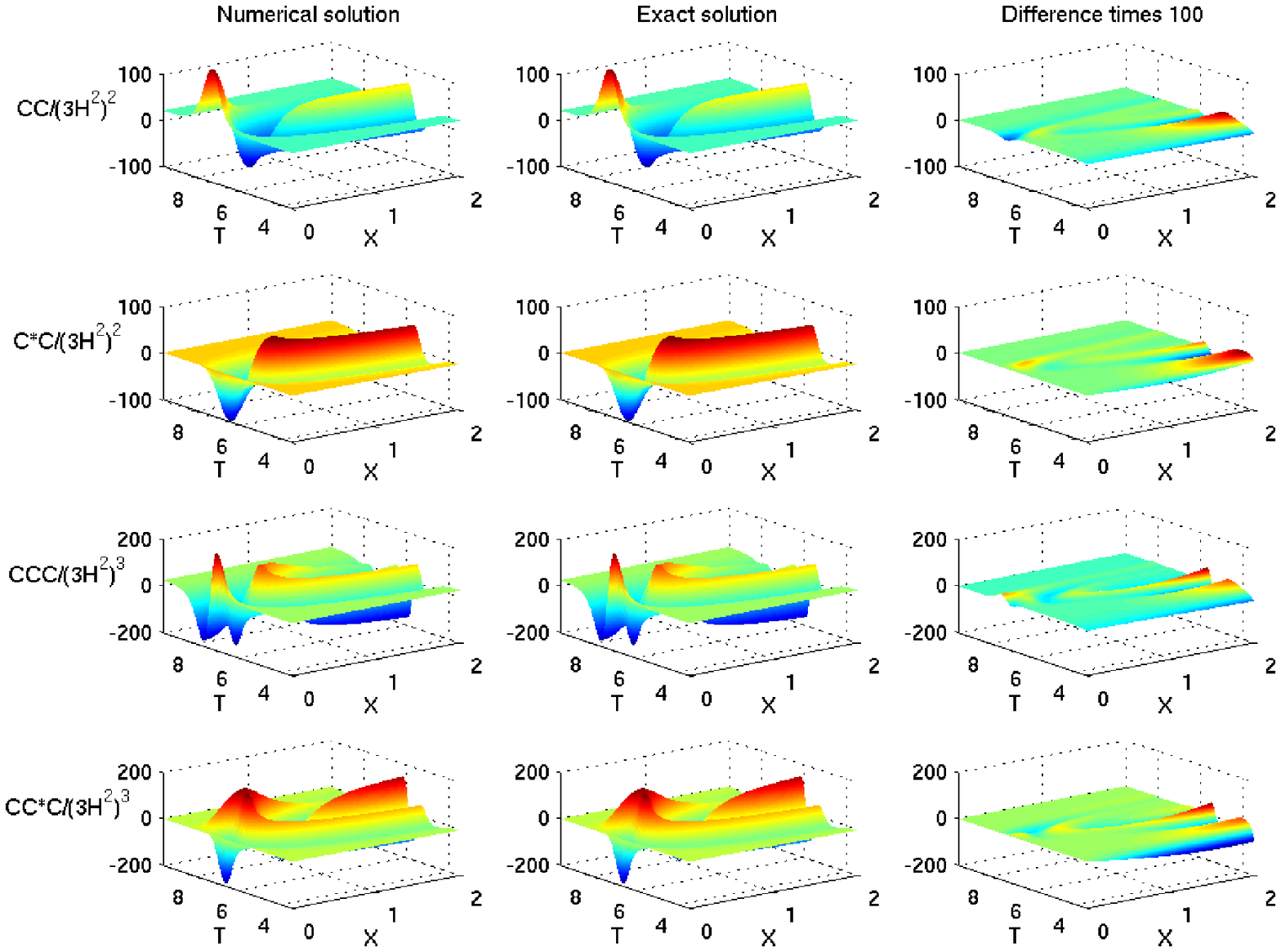}}
    \caption{\label{weyl_2}The Weyl scalar invariants (normalized by Hubble scalar)
    for the numerical solution, the matching exact spike solution (with $w=1.8329$, $\tau_0=6.6029$),
    and their difference, during the second spike recurrence over the time interval $[3.5,9.5]$.}
\end{figure*}

The simulation shows 
a perturbed spike solution recurs twice over the same Gowdy era.
For each of the two recurrences, the numerical solution is matched with
an exact spike solution. The difference between the numerical and exact spike solution
is computed in the four Weyl scalar invariants, and is observed to be smaller
in the second recurrence than in the first (see Figures~\ref{weyl_1} and \ref{weyl_2}). 
This suggests that the closer to the singularity, the closer the numerical solution 
gets to an exact spike solution.
This supports the conjecture that the exact spike solutions are attractors.

\begin{figure*}[t!]
    \resizebox{!}{4cm}{\includegraphics{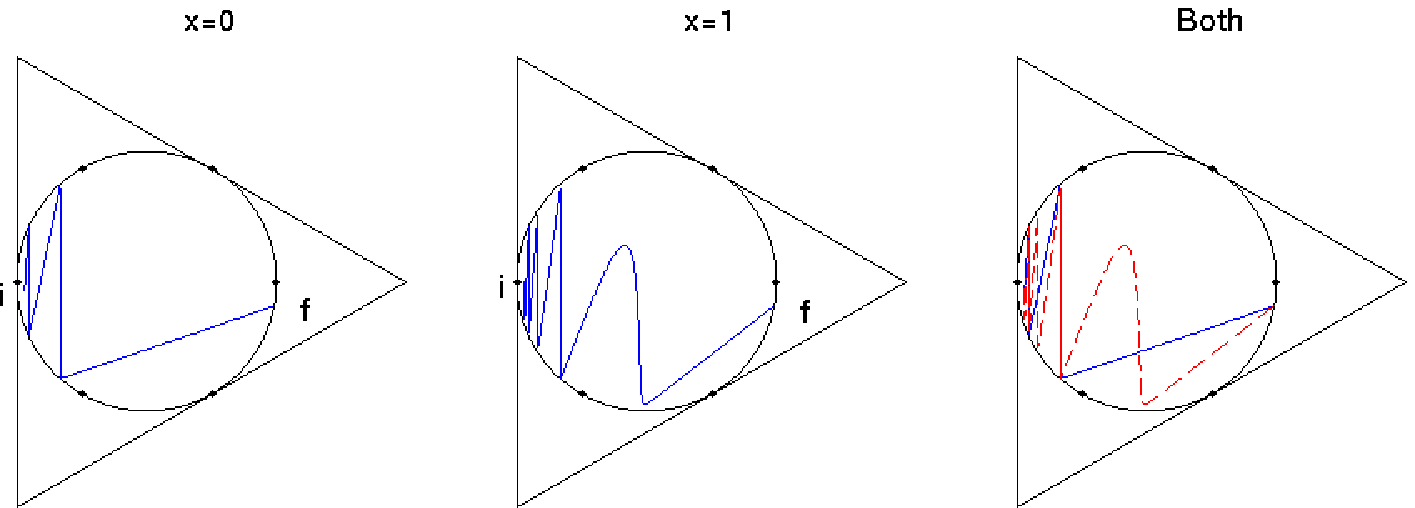}}
    \caption{\label{zoom_path}The orbit along the spike point $x=0$ and a nearby non-spike 
point $x=1$ projected onto the $(\Sph,\Smh)$ plane for the perturbed 
spike simulation. The initial (at $T=0$) and final (at $T=16$) points are 
marked by the letters i and f respectively.}
\end{figure*}

Figure~\ref{zoom_path} shows the orbit along the spike point $x=0$ (from the first simulation) and the 
worldline $x=1$ (from the second simulation) projected onto the $(\Sph,\Smh)$ plane in the state space of Hubble-normalized variables. 
It shows the orbits follow the expected paths as predicted from the spike solution and 
the $\Sc$ and $\Nm$ transition sets (see Figures 5 and 6 in \cite{L08}).
This subsection is similar to the work in \cite{GW03}, which was done in the Gowdy class ($\St=0$),
in the sense that the simulations here focus on what happens within one Gowdy era.
The approximate values for the $w$ parameter in \cite{L08} and corresponding $u$ parameter when near a Kasner point
for the $x=0$ orbit in Figure~\ref{zoom_path} are given below (rounded to 4 decimal points). 
Linking the Kasner epochs are alternating frame and spike transitions. 

\begin{gather}
	w \approx -7.8330 
	\stackrel{\text{frame}}{\longrightarrow} 7.5401
	\stackrel{\text{spike}}{\longrightarrow} -3.7592
	\stackrel{\text{frame}}{\longrightarrow} 3.8264
\notag\\
	\stackrel{\text{spike}}{\longrightarrow}  0.1674
\\
	u \approx                       3.4165
        \stackrel{u}{\longrightarrow}   3.2700 
	\stackrel{u-2}{\longrightarrow} 1.3796
	\stackrel{u}{\longrightarrow}   1.4132
\notag\\
	\stackrel{\frac{1}{u-1} -1}{\longrightarrow} 1.4021
\end{gather}
Note that Kasner epochs linked by frame transitions are not distinct physically.
One also observes that the numbers above do not follow the maps very closely, suggesting that the solution is not yet very close
to the generalized Mixmaster attractor. The difference is due to perturbation present in the initial data. Over time, the difference gradually 
decreases.
Also note that the map $u\rightarrow u-2$ has an adjustment algorithm when the new value is less than 1, namely
\be
	u\rightarrow 	\begin{cases}	u-2 & \text{if $u \geq 3$} \\
					\frac{1}{u-2} & \text{if $2 < u \leq 3$} \\
					\frac{1}{\frac{1}{u-1}-1} & \text{if $\frac{3}{2} \leq u < 2$} \\
					\frac{1}{u-1}-1 & \text{if $1 < u \leq \frac{3}{2}$.}
			\end{cases}
\ee

\subsection{Generic initial condition}

Having seen that perturbed spike initial data lead to recurring spikes within the same Gowdy era,
we now go further and ask whether a generic initial data also lead to recurring spikes,
and whether the recurrence continues in the next Gowdy era. To answer these questions,
we shall start with a generic initial data, and evolve the solution
through to the next Gowdy era.

We give an example of a generic initial condition for a true spike 
below.
\begin{gather}
\EEE=2,\qquad \St = (\St)_0 \e^{\int_0^x C_T/\EEE dx},
\\
\Sm = a_1 + a_2 x^2,\qquad \Nc = a_5 x,\qquad
\\
\Sc = a_3 + a_4 x^2,\qquad \Nm = a_6 x,
\end{gather}
with
\begin{align}
\int_0^x C_T/\EEE dx &= \frac14(-3 a_1 a_5 - \sqrt{3} a_5 + 3 a_3 a_6)x^2 
\notag\\
	&\quad	+\frac18(-3 a_2 a_5 + 3 a_4 a_6)x^4,
\end{align}
where $x=X+x_\text{zoom}$.
For example, we choose
\begin{gather}
	a_1=3.25,\qquad a_2=0.002,\qquad a_3=0.3,
\\      a_4=-0.001,\qquad a_5=0.04,\qquad a_6=-0.05,
\\
	(\St)_0 = 0.2,
\end{gather}
with $A=1$, $x_\text{zoom}=0$, 6401 grid points over the $X$-interval $[0,10]$, and
time interval $[0,40]$. Quadruple precision is used.
For comparison, another simulation with $x_\text{zoom}=1$, 12801 grid points over the $X$-interval $[-10,10]$ is used.
Beyond $T=40$, the solution gets too close to a Kasner point, and a higher numerical resolution is needed to maintain accuracy.

\begin{figure*}[t!]
    \resizebox{!}{4cm}{\includegraphics{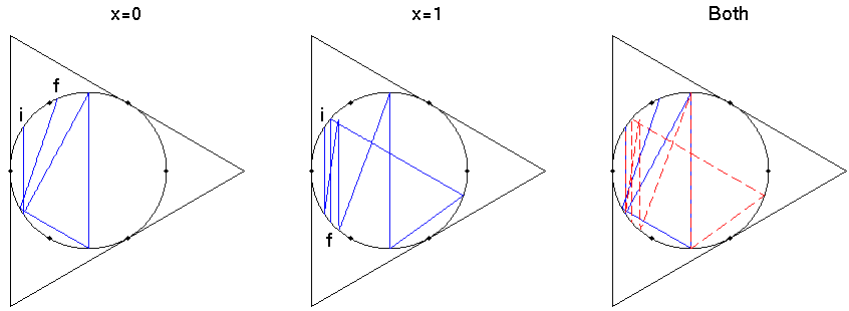}}
    \caption{\label{zoomic_path}The orbit along the spike point $x=0$ and a nearby non-spike 
point $x=1$ projected onto the $(\Sph,\Smh)$ plane for the generic 
initial data simulation. The initial (at $T=0$) and final (at $T=40$) points are
marked by the letters i and f respectively.}
\end{figure*}

Two recurrences of spike are observed, one in the same Gowdy era, and the other in the next (after a $\St$ transition).
Figure~\ref{zoomic_path} shows the orbits along $x=0,1$ passing close to
various identical points during two Gowdy eras. A difference in position of the final points
is observed, and is attributed to the lag between the two worldlines that becomes more pronounced
over time.
The approximate values for the $w$ and $u$ parameters when near a Kasner point (except the initial point)
for the $x=0$ orbit in Figure~\ref{zoomic_path} are given below (rounded to 4 decimal points).
\begin{gather}
        w \approx 5.7070
        \stackrel{\text{spike}}{\longrightarrow} -1.7114
	\stackrel{\text{frame}}{\longrightarrow} 1.7114
	\stackrel{\text{$\St$ frame}}{\longrightarrow} 6.6228
\notag\\
	\stackrel{\text{spike}}{\longrightarrow} -2.6228
\\
        u \approx 				  2.3535
	\stackrel{\text{$\frac{1}{u-2}$}}{\longrightarrow}  2.8114
        \stackrel{\text{$u$}}{\longrightarrow}    2.8114
	\stackrel{\text{$u$}}{\longrightarrow}    2.8114
\notag\\
	\stackrel{\text{$\frac{1}{u-2}$}}{\longrightarrow}  1.2324
\end{gather}
Observe that the numbers here follow the map much more closely in the later stage.

\begin{figure*}[t!]
    \resizebox{\linewidth}{!}{\includegraphics{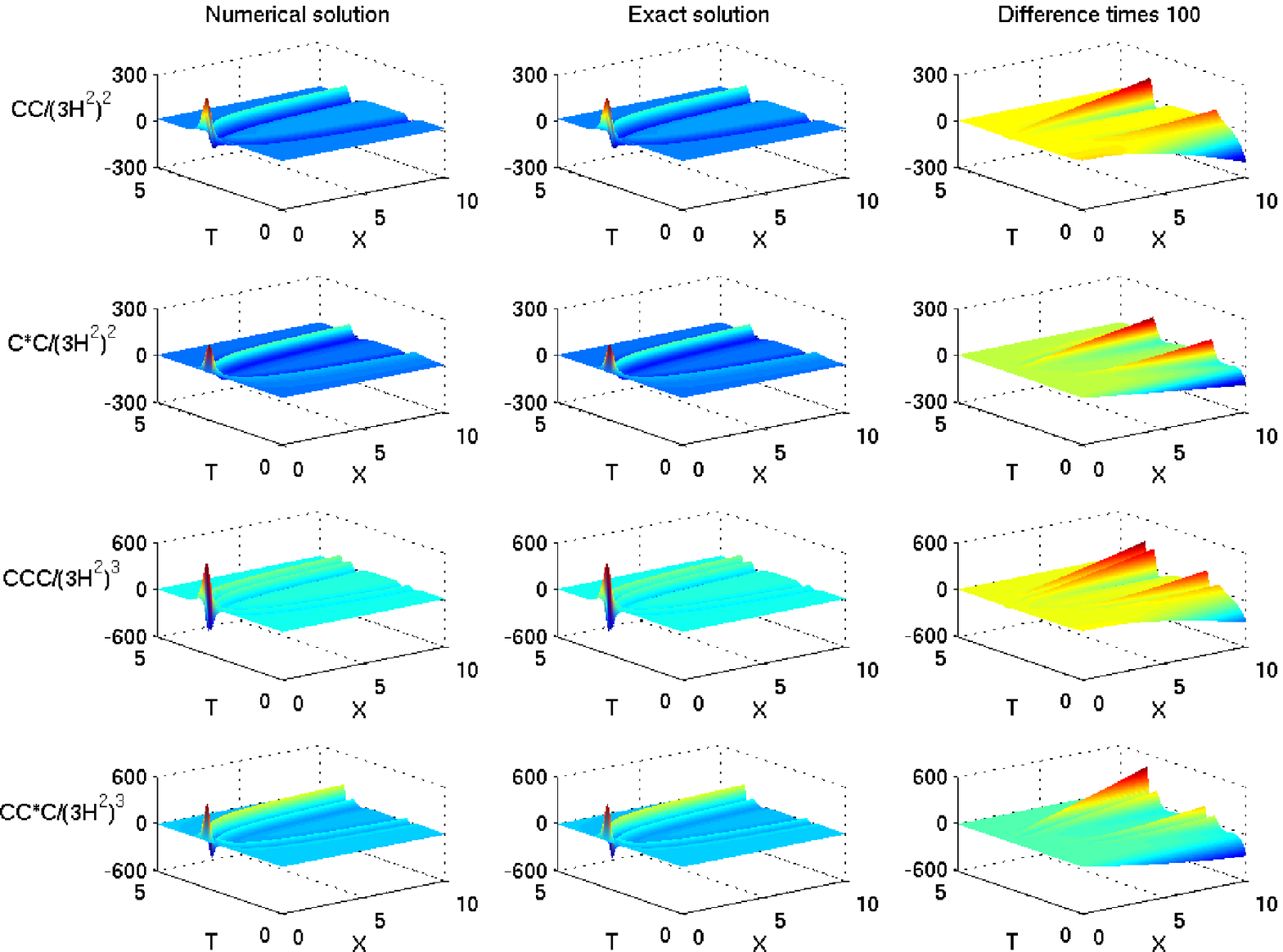}}
    \caption{\label{weyl_1g}The Weyl scalar invariants (normalized by Hubble scalar)
    for the numerical solution, the matching exact spike solution (with $w=-3.7114$, $\tau_0=2.6461$),
    and their difference, during the first spike recurrence over the time interval $[0,5]$.}
\end{figure*}

\begin{figure*}[t!]
    \resizebox{\linewidth}{!}{\includegraphics{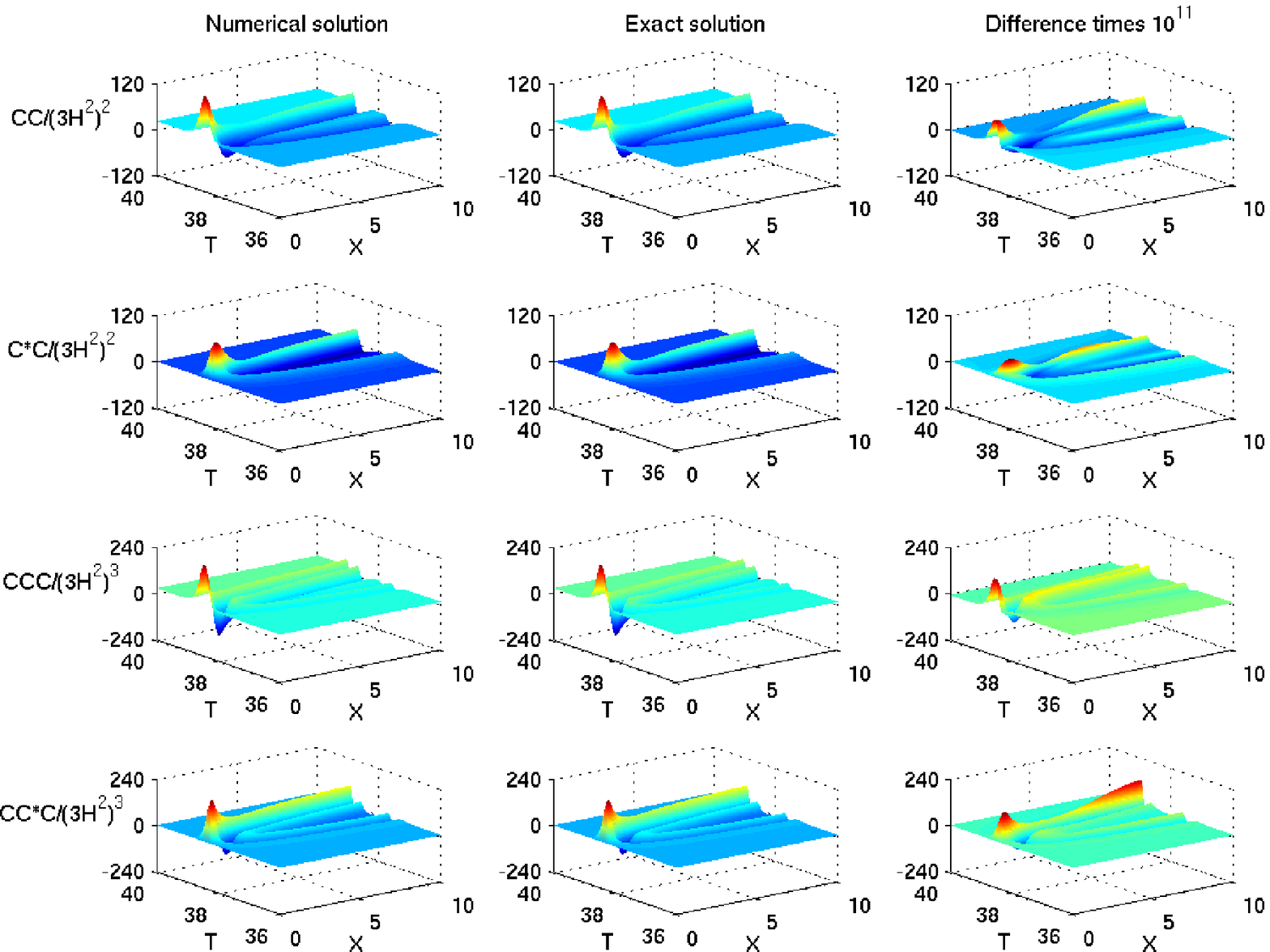}}
    \caption{\label{weyl_2g}The Weyl scalar invariants (normalized by Hubble scalar)
    for the numerical solution, the matching exact spike solution (with $w=-4.6228$, $\tau_0=38.1923$),
    and their difference, during the second spike recurrence over the time interval $[35,40]$.}
\end{figure*}

The numerical solution is matched with an exact spike solution 
and the Weyl scalars are plotted in Figures~\ref{weyl_1g} and \ref{weyl_2g}.
As in the previous subsection, matching improves with time (towards the singularity).
The remarkable improvement from the first to the second recurrence also suggest an exponential rate of
convergence to the exact spike solution. This provides a very strong evidence that
the spike solution is an attractor not only for perturbed spike initial data, but also for generic ones.
The exponential rate of convergence is also a curse for accurate numerical simulations,
as the need for numerical resolution also increases exponentially with time.

\subsection{Perturbed higher order spikes}

Having seen that the spike solution is an attractor, we now investigate
whether higher order spikes \cite[Section 5.5]{L08} are also attractors.
In this subsection we shall use perturbed second and third order spike solutions
as initial data, and see whether they recur as first, second or third order spikes.

The initial data for a perturbed second order spike solution 
at $\tau=0$ is given recursively in terms of the first order spike 
solution:
\begin{gather}
\EEE=2,
\\
\Sm = -(c{\Sm}_1+s{\Sc}_1)-\frac{1}{\sqrt{3}},\qquad
\Nc =   s{\Nm}_1-c{\Nc}_1,
\label{ic2b}
\\
\Sc = c{\Nm}_1+s{\Nc}_1,\qquad
\Nm = -s{\Sm}_1+c{\Sc}_1,
\label{ic2c}
\end{gather}
where $(\Sm,\Nc,\Sc,\Nm)_1$ are the perturbed first order spike solution 
in (\ref{ic1b})--(\ref{ic1c}). $c$ and $s$ are given by
\begin{gather}
c = \frac{f_1^2 -1}{f_1^2+1},\qquad
s = \frac{2f_1}{f_1^2+1},
\\
f_1 = \frac{1}{(wx)^2+1}\left[ -2w(1-2x^2)+\frac{Q_2}{2Q_0}\right],
\label{f1}
\end{gather}
where $x=X+x_\text{zoom}$.
$\St$ is specified by numerically evaluating the constraint
\be
\St = (\St)_0 \e^{\int_0^x C_T/\EEE dx}.
\ee

We perform a simulation with the parameters
\begin{gather}
	\epsilon = 10^{-3},\qquad
	w=9.5,\qquad
	A=1,
\\
	(\St)_0=10^{-3},\qquad
	Q_2=0,
\end{gather}
and $x_\text{zoom}=0$, $X \in [0,2]$ with 10001 grid points, and $T \in 
[0,12]$.

\begin{figure}[t!]
    \resizebox{!}{4cm}{\includegraphics{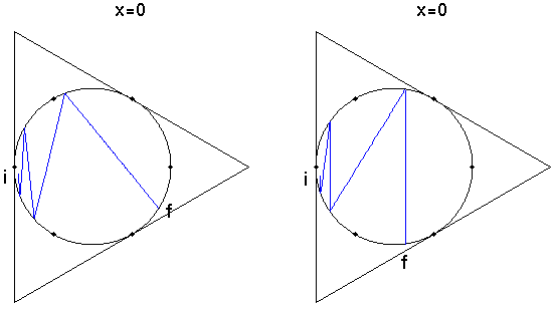}}
    \caption{\label{zoom2s3s_path}The orbit along the spike point $x=0$ projected onto the 
$(\Sph,\Smh)$ plane for the perturbed second (left) and third (right) 
order spike simulation. The orbits follow the false and the true spike 
orbits respectively. The initial (at $T=0$) and final (at $T=12$) points are
marked by the letters i and f respectively.}
\end{figure}

\begin{figure*}[t!]
    \resizebox{\linewidth}{!}{\includegraphics{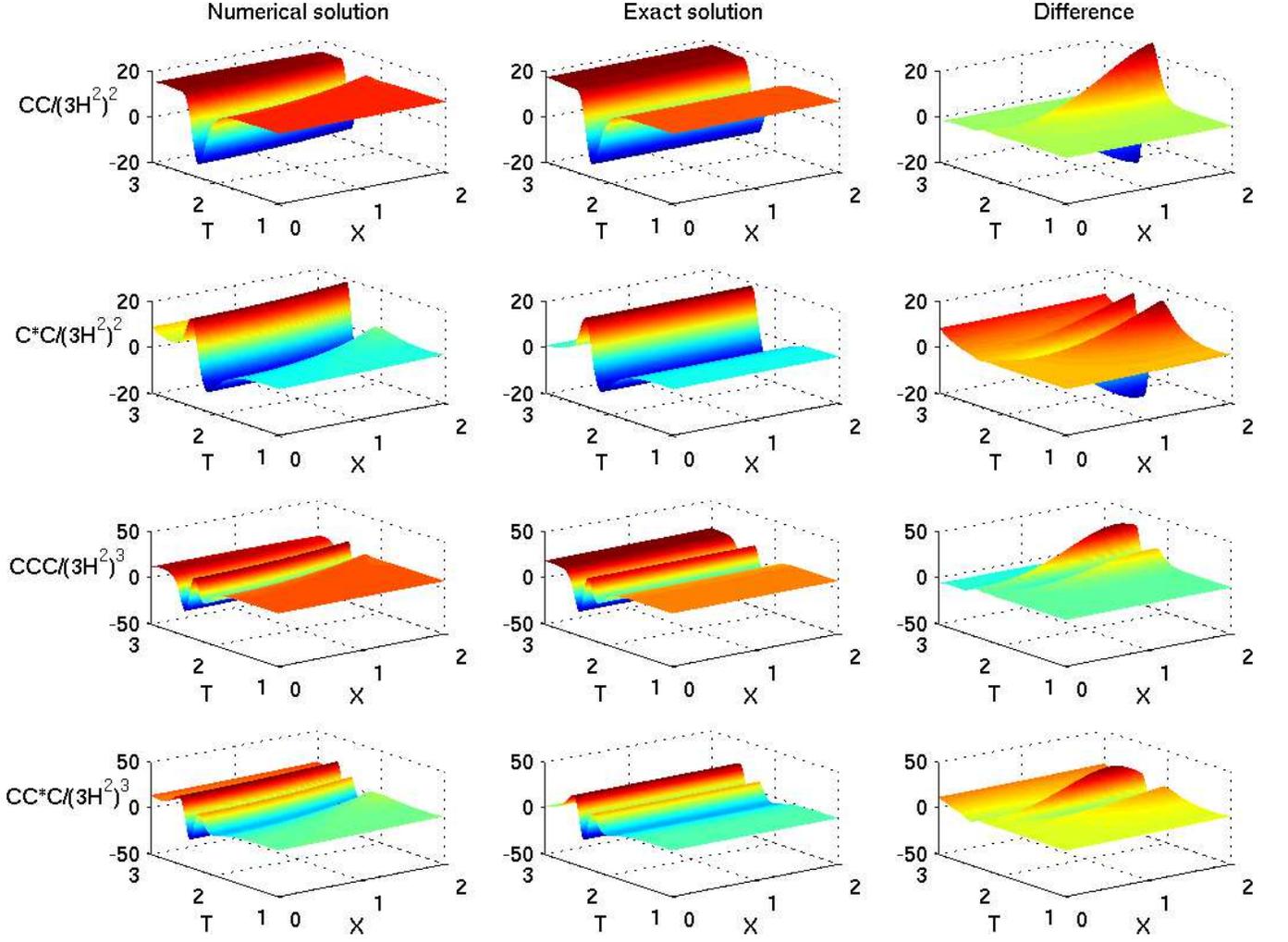}}
    \caption{\label{weyl_1_2s}The Weyl scalar invariants (normalized by Hubble scalar)
    for the numerical solution, the matching exact false spike solution (with $w=5.5005$, $\tau_0=2.3329$),
    and their difference, during the first false spike recurrence over the time interval $[1,3]$.}
\end{figure*}

\begin{figure*}[t!]
    \resizebox{\linewidth}{!}{\includegraphics{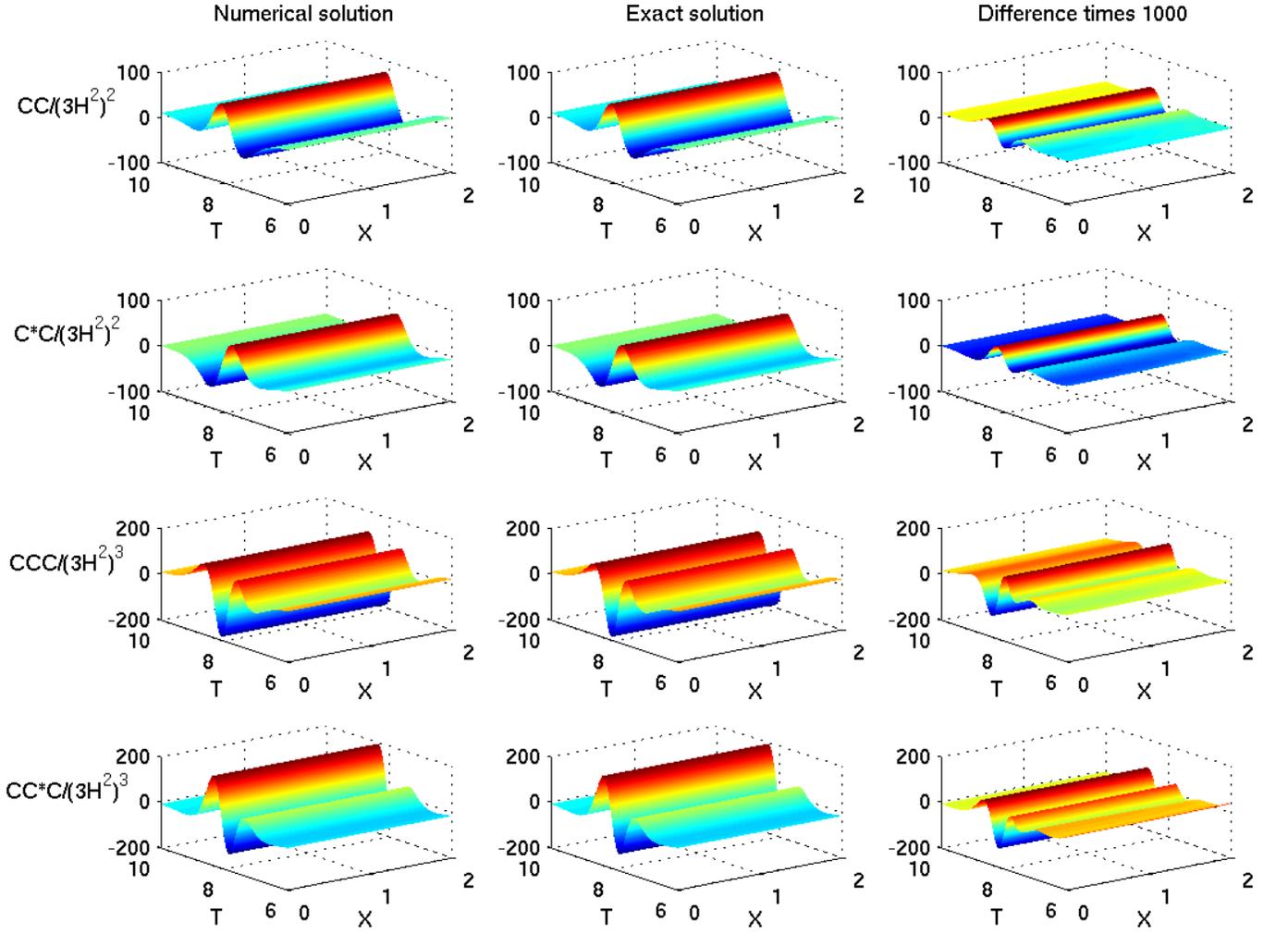}}
    \caption{\label{weyl_2_2s}The Weyl scalar invariants (normalized by Hubble scalar)
    for the numerical solution, the matching exact false spike solution (with $w=1.5006$, $\tau_0=7.6904$),
    and their difference, during the second false spike recurrence over the time interval $[6,10]$.}
\end{figure*}

\begin{figure*}[t!]
    \resizebox{\linewidth}{!}{\includegraphics{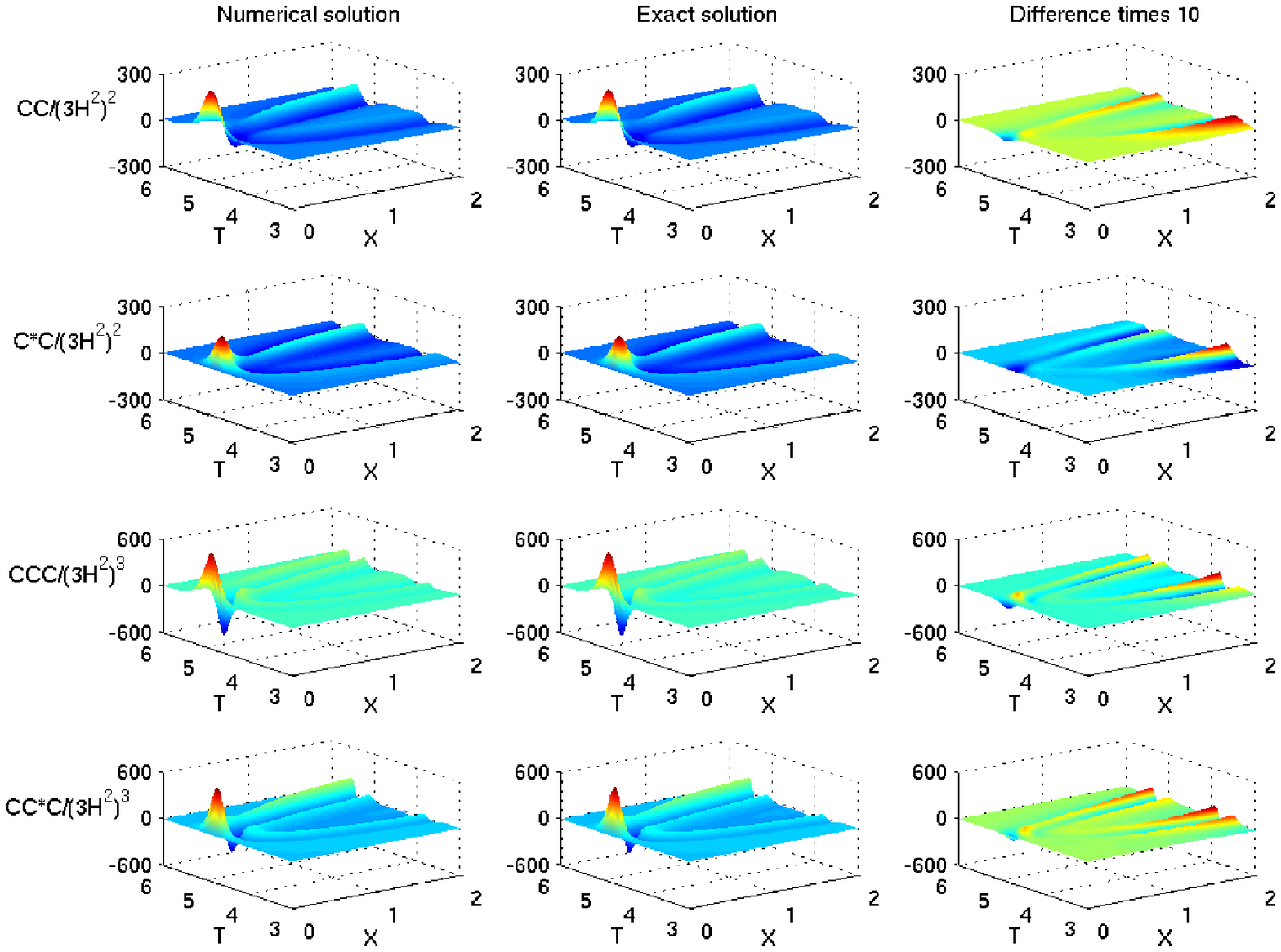}}
    \caption{\label{weyl_1_3s}The Weyl scalar invariants (normalized by Hubble scalar)
    for the numerical solution, the matching exact spike solution (with $w=-3.5002$, $\tau_0=4.4985$),
    and their difference, during the first spike recurrence over the time interval $[3,6]$.}
\end{figure*}

The initial data for a perturbed third order spike solution
at $\tau=0$ is also given recursively:
\begin{gather}
\EEE=2,
\\
\Sm = -(c{\Sm}_2+s{\Sc}_2)-\frac{1}{\sqrt{3}},\qquad
\Nc =   s{\Nm}_2-c{\Nc}_2,
\\
\Sc = c{\Nm}_2+s{\Nc}_2,\qquad
\Nm = -s{\Sm}_2+c{\Sc}_2,
\end{gather}
where $(\Sm,\Nc,\Sc,\Nm)_2$ are the perturbed second order spike solution
in (\ref{ic2b})--(\ref{ic2c}) above. $c$ and $s$ are given by
\begin{align}
c &= \frac{f_2^2 -1}{f_2^2+1},\qquad
s = \frac{2f_2}{f_2^2+1},
\\
	f_2 &= \frac{1}{[(wx)^2+1][f_1^2+1]}
\notag\\	
	&\quad
	\times\Bigg[
	-\frac{1}{3}x\Big(3[(wx)^2+1][(f_1^2-1)w-4f_1]
\notag\\
	&\qquad
	+4w(w^2+2)x^2\Big)
	+\frac{Q_3}{2Q_0}
	\Bigg],
\end{align}
where $f_1$ is given in (\ref{f1}), and $x=X+x_\text{zoom}$.
$\St$ is specified by numerically evaluating the constraint
\be
\St = (\St)_0 \e^{\int_0^x C_T/\EEE dx}.
\ee

We perform a simulation with the parameters
\begin{gather}
	\epsilon=10^{-3},\qquad
	w=9.5,\qquad
	A=1,
\\
	(\St)_0=10^{-3},\qquad
	Q_2=0,\qquad
	Q_3=0,
\end{gather}
and $x_\text{zoom}=0$, $X \in [0,2]$ with 10001 grid points, and 
$T \in [0,12]$.

The center or inner part of a perturbed second order spike evolves into a false first order spike,
as suggested by Figures~\ref{weyl_1_2s} and \ref{weyl_2_2s}.
False spikes are merely a spiky representation of the vacuum Bianchi type II solution.
The center or inner part of a perturbed third order spike evolves into a true first order spike,
as suggested by Figure~\ref{weyl_1_3s}.
The outer parts of the perturbed spikes move beyond the domain of simulation and
are suspected to evolve into first order spikes, with a moving spike point.
A global numerical scheme might be needed to follow their evolution, 
but one with enough numerical resolution would take months to run, which is impractical.
Figure~\ref{zoom2s3s_path} shows that the orbit for a perturbed second 
(third) order spike later follows the predicted orbit for the false (true) 
first order spike. 
This suggests that higher order spikes break into separate first order spikes, 
and therefore are not attractor themselves.

The approximate values for the $w$ and $u$ parameters when near a Kasner point (except the first point)
for the $x=0$ orbits in Figure~\ref{zoom2s3s_path} are given below (rounded to 4 decimal points).
For the left figure, two false spike transitions and ($\Sc$) curvature transition link the Kasner epochs.
\begin{gather}
        w \approx -6.5005
	\stackrel{\text{false spike}}{\longrightarrow} 4.4161
	\stackrel{\text{curvature}}{\longrightarrow} 2.4980
\notag\\
	\stackrel{\text{false spike}}{\longrightarrow} 0.5003
\\
        u \approx  2.7502
	\stackrel{\text{$u-1$}}{\longrightarrow} 1.7081
	\stackrel{\text{$\frac{1}{u-1}$}}{\longrightarrow} 1.3351
	\stackrel{\text{$\frac{1}{u-1}$}}{\longrightarrow} 3.0024
\end{gather}
Recall that false spike transitions and curvature transition are physically the same.

For the right figure, two frame transitions and a spike transition link the Kasner epochs.
\begin{gather}
        w \approx -5.5004
	\stackrel{\text{frame}}{\longrightarrow} 5.4999
	\stackrel{\text{spike}}{\longrightarrow} -1.4979
	\stackrel{\text{frame}}{\longrightarrow} 1.5002
\\
        u \approx  		                 2.2502
	\stackrel{\text{$u$}}{\longrightarrow}   2.2500 
	\stackrel{\text{$\frac{1}{u-2}$}}{\longrightarrow} 4.0169
        \stackrel{\text{$u$}}{\longrightarrow}   3.9984
\end{gather}

\section{Conclusion}

We have found numerical evidence (from both perturbed solutions and 
generic initial data) that the spike solution is part of the generalized 
Mixmaster attractor.
We have found that the second and third order spikes are not part of the 
attractor, and conjecture that all higher order spikes are not part of
the attractor.

We summarize the above conjectures as follows:
\begin{enumerate}
\item
Spike transitions are a new type of oscillation on approach to the singularity,
with each transition approximated by a spike solution.
A spike transition has a map of $u \rightarrow u-2$ and
is different from the previously known Mixmaster oscillation, which has a map of 
$u \rightarrow u-1$. It occurs in a causal neighborhood of special 2D surfaces of
worldlines in generic spacetimes.

\item
Higher order spike transitions (with maps $u \rightarrow u-3$, etc) split into
first order spike transitions and so are not general. i.e. the generic behavior
towards singularity is either $u \rightarrow u-1$ or $u \rightarrow u-2$.
\end{enumerate}

We have used symmetric data in order to hold the spike point fixed, so 
that we can zoom in on it. We believe that for non-symmetric data, in 
which the spike point can move (by a little when the spike is active, and 
sometimes by a lot when the spike is smoothed out), the above conclusion 
should also hold. This remains to be confirmed numerically.
At present we do not know how to zoom in on a moving spike point.

What remains unanswered is the following.
Because we simulate only the neighborhood of a spike, we do not know 
what happens outside this domain. We also do not know what happens to new 
spike points that are created and move out of the domain, how they 
interact with other spike points or false spike points. 
Existing numerical simulations from the global view suffers from expensive 
resources needed to resolve spikes, which severely limit the length of 
simulation. We envision a new way to simulate spikes, by combining the 
zoom-in view with the global view. 
The biggest benefit of such a combination is much longer simulations.
The zoom-in view can also provide boundary conditions, so that the 
assumption of spatial periodicity can be dropped.
Implementing the combination will be challenging.

\appendix
\section{Matching with explicit solutions}\label{App_A}

For the purpose of matching the numerical solutions with explicit solutions,
we will need the explicit spike solutions with generic time and space constants.
To restore these constant, perform the transformation
\be
	\tau \rightarrow \tau-\tau_0,\quad
	x \rightarrow \frac{2}{(\EEE)_0}(x-x_0).
\ee
The expression of the metric, the governing equations, and the solutions will
change accordingly.
In particular, $\EEE$ is now given by
\be
	\EEE = (\EEE)_0 e^{\tau-\tau_0}.
\ee
The spike solution is now given by
\begin{align}
        P &= 2(\tau-\tau_0) + \ln[\s]-\ln[f^2+1] 
\notag\\
	&\qquad
	-\ln(2Q_0)
\\
        Q &= -Q_0 w [ 2(w\t-1)(\tfrac{2}{(\EEE)_0}(x-x_0))^2
\notag\\
        &\qquad
	+ \e^{-2(\tau-\tau_0)} ]
	+ Q_2
\\
        \lambda &=-4\ln[\s] + 2\ln[f^2+1] 
\notag\\
        &\qquad
	-(w^2+4)(\tau-\tau_0)
                +\lambda_2,
\end{align}
where
\be
	f = w\e^{\tau-\tau_0}\s \frac{2}{(\EEE)_0}(x-x_0)
\ee
is the the factor $Q\e^P$ for the vacuum Bianchi type II solution.
Correspondingly, the $\beta$-normalized variables become
\begin{align}
\Sm &= {\frac {-1} {\sqrt 3}} \bigl ( 1 + {\frac {{f^2} - 1} {{f^2}+1}}
[ w \tanh (w(\tau-\tau_0) ) - 1 ] \bigr )
\\
\Nc &= {\frac {2 f} {{f^2}+1}} {\frac w {\sqrt 3}} \sech (w (\tau-\tau_0))
\\
\Sc &= {\frac {{f^2} - 1} {{f^2}+1}} {\frac w {\sqrt 3}} \sech (w (\tau-\tau_0))
\\
\Nm &=  {\frac {2 f} {{f^2}+1}} {\frac 1 {\sqrt 3}} [ 1 - w \tanh (w(\tau-\tau_0) )].
\end{align}
The other solutions are similarly restored.

We take this opportunity to correct errors in \cite{L08}:
the third minus sign in Equation (28) should be a plus sign, and the factor 4
in Equation (34) should not be there.

In order to match with an explicit spike solution, we will need to guess
the value of the parameter $w$. This can be done in two ways.
The first way is to choose a predetermined value, the second is
to obtain a guess from the numerical solution. To do so we compute the expression
\be
	\text{arcsinh}\left(\frac{\Sm+\tfrac{2}{\sqrt{3}}}{\Sc}\right),
\ee
along $X=0$, which equals
\be
	w(\tau-\tau_0)
\ee
for the spike solution.
The value $\tau_0$ is then obtained through interpolation.

We then compute the expression
\be
	\left(\Sm+\tfrac{2}{\sqrt{3}}\right)^2+\Sc^2
\ee
along $X=0$. For the spike solution this expression equals $w^2/3$.
In practise the numerical solution will give an close-to-constant time function
of $w$, from which we choose one value. For example we can take the maximum value
of this time function. 

\section{Obtaining the BKL parameter $u$ for the Kasner epochs}\label{App_B}

Matching the Kasner epochs with Kasner solutions is straightforward.
Recall from equation (18) of \cite{L08} that for a Kasner solution
\be
	\Sm = -\frac{w}{\sqrt{3}},
\ee
where $w$ is a constant.
One then obtains the local maximum and minimum values for $\Sm$ along a worldline
and convert them to $w$. Then one computes the BKL parameter $u$ from $w$ using the following
formula. 
\be
	u = \begin{cases}
		\frac{|w|-1}{2} & |w| \geq 3,
\\
		\frac{2}{|w|-1} & 1 < |w| \leq 3,
\\
		\frac{1+|w|}{1-|w|} & 0 \leq w < 1.
	\end{cases}
\ee

\section{The Weyl scalar invariants}\label{App_C}

The orthonormal frame components $C_{abcd}$ of the Weyl tensor
can be conveniently expressed in terms of the electric and magnetic
components $E_{\alpha\beta}$ and $H_{\alpha\beta}$ \cite{vEU97}:
\begin{gather}
        C_{\alpha0\beta0}=E_{\alpha\beta},\qquad
        C_{\alpha\beta\gamma\delta}=
        -\epsilon^\mu{}_{\alpha\beta}\epsilon^\nu_{\gamma\delta}E_{\mu\nu},
\\
        C_{\alpha\beta\gamma0}=\epsilon^\mu{}_{\alpha\beta}H_{\gamma\mu},
\end{gather}
which are then normalized by $3\beta^2$:
\be
        \E_{\alpha\beta}=\frac{1}{3\beta^2}E_{\alpha\beta},\qquad
        \H_{\alpha\beta}=\frac{1}{3\beta^2}H_{\alpha\beta},
\ee
and further decomposed as follows:
\be
        \mathcal{E}_{\alpha \beta}
        = \left(\begin{matrix}
        -2\Ep & \sqrt{3} \E_3 & \sqrt{3} \E_2 \\
        \sqrt{3} \E_3 & \Ep + \sqrt{3} \Em & \sqrt{3} \Ec \\
        \sqrt{3} \E_2 & \sqrt{3} \Ec & \Ep - \sqrt{3} \Em
        \end{matrix}\right)
\ee
and similarly for $\H_{\alpha\beta}$.
The components are given by
\begin{align}
\Ep &= \tfrac13\Sp-\tfrac13(\Sm^2+\Sc^2)+\tfrac23(\Nm^2+\Nc^2)+\tfrac16\St^2
\\
\Em &= \tfrac13(1-3\Sp)\Sm+\tfrac23\Np\Nm
\notag\\
	&\qquad
	+\tfrac13(\di-r)\Nc+\tfrac1{2\sqrt{3}}\St^2
\\
\Ec &= \tfrac13(1-3\Sp)\Sc+\tfrac23\Np\Nc
\notag\\
        &\qquad
	-\tfrac13(\di-r)\Nm
\\
\E_3 &= -\tfrac1{\sqrt{3}}\Sc\St
\\
\E_2 &= \tfrac13(1+\sqrt{3})\Sm)\St
\\
\Hp &= -\Nm\Sm-\Nc\Sm
\\
\Hm &= -\Sp\Nm-\tfrac23\Np\Sm-\tfrac13(\di-r)\Sc
\\
\Hc &= -\Sp\Nc-\tfrac23\Np\Sc+\tfrac13(\di-r)\Sm
\\
\H_3 &= -\tfrac1{\sqrt{3}}\Nc\St
\\
\H_2 &= \tfrac1{\sqrt{3}}\Nm\St
\end{align}
where $\Np=\sqrt{3}\Nm$, $r=-3(\Nc\Sm-\Nm\Sc)$.
The four Weyl scalar invariants are computed as follows:
\begin{align}
        C_{abcd}C^{abcd}
&= 8 (E_{\alpha\beta}E^{\alpha\beta} - H_{\alpha\beta}H^{\alpha\beta})
\\
        C_{abcd}{}^*C^{abcd}
        &= 16 E_{\alpha\beta} H^{\alpha\beta}
\\
        C_{ab}{}^{cd}C_{cd}{}^{ef}C_{ef}{}^{ab}
        &= -16 (E_\alpha{}^\beta E_\beta{}^\gamma E_\gamma{}^\alpha
        -3 E_\alpha{}^\beta H_\beta{}^\gamma H_\gamma{}^\alpha)
\\
        C_{ab}{}^{cd}C_{cd}{}^{ef} {}^*C_{ef}{}^{ab}
        &= 16 (H_\alpha{}^\beta H_\beta{}^\gamma H_\gamma{}^\alpha
        -3 E_\alpha{}^\beta E_\beta{}^\gamma H_\gamma{}^\alpha),
\end{align}
where ${}^*C_{abcd}=\tfrac12\eta_{ab}{}^{ef}C_{efcd}$,
and $\eta^{abcd}$ is the totally antisymmetric permutation tensor, with
$\eta^{0123}=1$.

The drawback of plotting the Weyl scalars for spikes is that the blow-up
of the Weyl scalars towards the singularity makes the spiky structures invisible.
For example, in Figure 8 of \cite{L08}, level curves have to be plotted to make
the structure visible.
In this paper we plot Hubble-normalized Weyl scalars so that spiky structures are clearly visible.
The Weyl scalars are normalized as follows:
\begin{gather}
	\frac{C_{abcd}C^{abcd}}{(3H^2)^2},\qquad
	\frac{C_{abcd}{}^*C^{abcd}}{(3H^2)^2},
\\
	\frac{C_{ab}{}^{cd}C_{cd}{}^{ef}C_{ef}{}^{ab}}{(3H^2)^3},\qquad
	\frac{C_{ab}{}^{cd}C_{cd}{}^{ef} {}^*C_{ef}{}^{ab}}{(3H^2)^3}.
\end{gather}

\begin{acknowledgments}
LA is partially supported by the NSF, grant no. DMS 0707306.
DG is partially supported by the NSF, grant no. PHY 0456655.
FP is partially supported by the NSF, grant no. PHY 0745779, and the Alfred P. Sloan Foundation.
WCL and LA thank the Mittag-Leffler Institute for hospitality during part of the work on this paper.
\end{acknowledgments}

\end{document}